\documentclass[times,twocolumn,final]{elsarticle}

\usepackage{framed,multirow}

\usepackage{amssymb}
\usepackage{amsmath}
\usepackage{latexsym}
\usepackage{algorithm}

\usepackage{ulem}

\usepackage{siunitx}

\usepackage{booktabs}
\usepackage{subfig}
\usepackage{gensymb}

\usepackage{algorithmicx}
\usepackage{algpseudocode}

\usepackage[switch]{lineno}

\newcommand{\argmin}{\operatornamewithlimits{argmin}}

\journal{Medical Image Analysis}

\begin{document}

\begin{frontmatter}

\title{Deep learning reconstruction of digital breast tomosynthesis images for accurate breast density and patient-specific radiation dose estimation}

\author[a1,a2]{Jonas {Teuwen}\fnref{fn1}}
\author[a1,a2]{Nikita {Moriakov}\fnref{fn1}}
\author[a1]{Christian {Fedon}}
\author[a1]{Marco {Caballo}}
\author[a5]{Ingrid {Reiser}}
\author[a6,a7]{Pedrag {Bakic}}
\author[a3]{Eloy {Garc\'ia}}
\author[a4]{Oliver {Diaz}}
\author[a1]{Koen {Michielsen}}

\author[a1,a8]{Ioannis Sechopoulos}
\cortext[cor1]{Corresponding author: Ioannis Sechopoulos} 
\fntext[fn1]{Equal contribution.}
\ead{ioannis.sechopoulos@radboudumc.nl}

\address[a1]{Department of Medical Imaging, Radboud University Medical Center, the Netherlands}
\address[a2]{Department of Radiation Oncology, Netherlands Cancer Institute, the Netherlands}
\address[a3]{Vall d'Hebron Institute of Oncology, VHIO, Spain}
\address[a4]{Department of Mathematics and Computer Science, University of Barcelona, Spain}
\address[a5]{Department of Radiology, The University of Chicago, USA}
\address[a6]{Department of Radiology, University of Pennsylvania, USA}
\address[a7]{Department of Translational Medicine, Lund University, Sweden}
\address[a8]{Dutch Expert Centre for Screening (LRCB), the Netherlands}

\begin{abstract}The two-dimensional nature of mammography makes estimation of the overall breast density challenging, and estimation of the true patient-specific radiation dose impossible. Digital breast tomosynthesis (DBT), a pseudo-3D technique, is now commonly used in breast cancer screening and diagnostics. Still, the severely limited 3rd dimension information in DBT has not been used, until now, to estimate the true breast density or the patient-specific dose. This study proposes a reconstruction algorithm for DBT based on deep learning specifically optimized for these tasks. The algorithm, which we name DBToR, is based on unrolling a proximal-dual optimization method. The proximal operators are replaced with convolutional neural networks and prior knowledge is included in the model. This extends previous work on a deep learning-based reconstruction model by providing both the primal and the dual blocks with breast thickness information, which is available in DBT. Training and testing of the model were performed using virtual patient phantoms from two different sources. Reconstruction performance, and accuracy in estimation of breast density and radiation dose, were estimated, showing high accuracy (density $< \pm3\%$; dose $< \pm 20\%$) without bias, significantly improving on the current state-of-the-art. This work also lays the groundwork for developing a deep learning-based reconstruction algorithm for the task of image interpretation by radiologists.
\end{abstract}

\end{frontmatter}
\section{Introduction}
Breast cancer screening with mammography has proven effective in reducing breast cancer-related mortality \citep{Plevritis2018}. However, since mammography is a 2D imaging modality, it results in the projection of the internal tissue of the breast onto a single plane, yielding tissue superposition. This results in a lower sensitivity and specificity, especially with dense breasts. 

\subsection{Digital breast tomosynthesis}
Digital breast tomosynthesis (DBT) has been introduced over the last two decades to decrease the impact of tissue superposition in mammography, by providing limited depth information, resulting in improved detection and diagnosis performance \citep{malmostudy,Zuley2013}. Hence, DBT is rapidly replacing digital mammography as the primary x-ray technique for breast imaging \citep{Niklason1997}. DBT imaging consists of acquiring several low-dose planar x-ray projections over a limited angle. These projections are then used to reconstruct a pseudo-3D volume, albeit with very limited vertical spatial resolution. A secondary benefit of the introduction of DBT for widespread use for screening offers the opportunity, for the first time, to obtain accurate estimates of the breast density and, subsequently, of the fibroglandular dose.

\subsection{Breast density}
In mammography, the lack of information on the true tissue distribution in the vertical (x-ray source to detector) direction limits the ability to accurately estimate the breast density (i.e., the proportion of the breast that consists of fibroglandular tissue). Up to now, estimates of breast density from mammography necessitated the use of models of the image acquisition process and assumptions and simplifications regarding tissue distribution. Although these models have been characterized for consistency and precision, their accuracy is unknown. Estimating breast density is of intense interest because it is an important factor for both masking and breast cancer risk \citep{McCormack2006}. As a result of its significance, breast density reporting is now mandated by law in several states in the USA. This makes it especially important to use methods that provide objective and quantitative breast density estimates. 
The feasibility of accurate breast density estimation from DBT images has been recognized before. However, given the very poor spatial resolution in the vertical direction obtained with all current DBT reconstruction algorithms, the localization of the fibroglandular tissue in DBT images has proven extremely challenging. Previous efforts to achieve this have not resulted in improvements over the model-based dose estimates obtained from 2D imaging \citep{Geeraert2014}, or have involved algorithms that require a lot of manual input, making them challenging to implement clinically \citep{Fornvik2018}.

\subsection{Radiation dose}
In breast imaging, the dose of interest is only that to the fibroglandular tissue, since this is the tissue most at risk of developing breast cancer \citep{Dance2016}, and to determine this dose it is necessary to know its vertical location within the breast. Therefore, currently, all breast dosimetry is based on dose estimates using a model breast, which does not reflect the dose deposited in the actual patient breast. It has been shown that the use of a model breast results in an average overestimation of the true patient breast of 30\% and that this error can be as high as 120\%, if not more \citep{Dance2005, Sechopoulos2012, Hernandez2015}. Even if a more accurate, unbiased model of the average breast is developed, an effort that is currently ongoing \citep{Arana2020}, the over 100\% error in patient-specific dose estimates using a population-wide model will not be ameliorated. To obtain accurate radiation dose estimates, the actual amount and position of the fibroglandular tissue in the individual patient's breast needs to be considered. However, due to its complete lack of information on the vertical position of tissues, this is impossible to achieve with mammography. Only with the introduction of digital breast tomosynthesis, is it now feasible to gather this knowledge for each imaged breast. 

Therefore, in this work, we propose a new approach to DBT reconstruction, based on our earlier work \citep{Moriakov}, using deep learning methods, that results in the 3D representation of the imaged breast optimized for estimation of the true distribution of the fibroglandular tissue. This in turn allows for accurate estimation of both the breast density and the radiation dose imparted on it, improving significantly upon the state-of-the-art.

\subsection{DBT reconstruction}
Both FBP-based and iterative methods are in clinical use to reconstruct breast tomosynthesis images, and both methods result in severely limited resolution in the direction perpendicular to the detector plane \citep{Sechopoulos2012,Vedantham2015}. Despite this, most research is focused on improving image quality in the high-resolution planes parallel to the detector plane since only this direction is examined by radiologists. In particular, the choice of the reconstruction algorithm and regularization parameters can greatly influence the reconstruction quality, as was shown in previous work \citep{rodriguez2017,Michielsen2016}.
Approaches that try to improve image quality in the vertical direction typically require strong prior knowledge to sufficiently constrain the inverse problem. In industrial settings this is feasible if the scanned object contains a small number of known materials  by applying a discrete tomography method such as poly-DART \citep{Six2019}.  A different approach recently proposed by \cite{Zhang2021} obtained promising results under the assumption that the true total variation in each direction of the scanned object is known.

\subsection{Deep learning for reconstruction}
A recent and very promising development in medical imaging reconstruction uses methods that rely on deep learning \citep{schoenlieb2019}. The goal of this paper is to show the potential of such methods for the problem of DBT reconstruction, using the quantitative estimation of breast density and radiation dose as the target application. Our method combines a deep learning network with an inductive bias given by the forward and backward models (and therefore considering part of the physics processes involved in image acquisition). This is in contrast to other established methods that postprocess initial reconstructions with a deep learning network to improve image quality \citep{Kang2017,Jin2017}. To build this algorithm, we extended our previous results on DBT reconstruction \citep{Moriakov} and investigate a data-driven reconstruction algorithm called Deep Breast Tomographic Reconstruction (DBToR), where the reconstruction is computed from projection data with a deep neural network. To train the model, and to test the performance of DBToR for the tasks of breast density and radiation dose estimation, we used dedicated breast CT images, where the tissue distribution is known, and use a finite-element model to simulate the change in the tissue distribution when under compression in DBT. In addition to these images, we also evaluate the model on simulated breast phantoms.

\subsection{Our contribution}
In this work, we show: (i) the feasibility of reconstruction of DBT using deep learning, having developed a novel deep learning-based model, to reconstruct DBT; (ii) that the resulting reconstructions have greatly improved vertical resolution compared to state-of-the-art analytical and iterative reconstruction methods; (iii) that the proposed reconstruction method is able to provide accurate breast density estimates; and (iv) that the dense tissue location information results in accurate patient-specific dosimetric estimates.

This paper is organized as follows: Section \ref{sec:inv} sets the context, describing DBT reconstruction as an inverse problem, and presents the architecture and model. Section \ref{sec:materials} describes the dataset and methods used for the dosimetric evaluation. We follow in Section \ref{sec:results} with the results obtained with the DBToR algorithm, and providing several comparisons against existing algorithms. Finally, we conclude in Section \ref{sec:conclusion}.

\section{Deep learning-based reconstruction}\label{sec:inv}

Before we describe our model in more detail, we give a brief overview on how to formulate the reconstruction problem as an inverse problem.

\subsection{Inverse problems and DBT reconstruction}
DBT reconstruction can be formulated as an inverse problem. In a mathematically simplified setting this means that given an object to be imaged $x \in X$ and measured projection data $y \in Y$, we have

\begin{equation}\label{eq:inv}
y = \mathcal{P} x + \eta
\end{equation}

\noindent where $\mathcal{P}: X \to Y$ is the forward, or projection, operator, that models how the object $x$ gives rise to the projection $\mathcal{P}x$ in the absence of noise, and $\eta$ is a $Y$-valued random variable modeling the noise component of the measurements. The forward model used is given by

\begin{equation}
    \label{eq:forward-model}
	y_i (\mathbf{x}) = b_i \exp\Bigl(- \sum_j l_{ij} x_j\Bigr),
\end{equation}

\noindent where $y_i$ is the projection data, $b_i$~the number of x-ray photons emitted towards detector pixel~$i$ and $l_{ij}$~the length of the intersection between voxel~$j$ and the line between the source and detector pixel~$i$. The linear attenuation in voxel~$j$ is denoted by $x_j$. 

The noise vector $\eta$ is typically assumed to be additive Gaussian, which is a good approximation for high photon counts, which are common in transmission imaging. In our work, we do not require an explicit noise model for inversion, and assume a more realistic noise model given by Poisson noise in Section \ref{sec:materials} for the simulations of the projection data.

The goal of reconstruction is to retrieve the object $x$ from measured (and noise-corrupted) projection data $y$. Inversion of the operator $\mathcal{P}$ is an ill-posed problem, and we will consider the estimation of $x$ from a Bayesian perspective and consider $X$ and $Y$ as probability spaces, therefore identifying in this work the spaces $X$ and $Y$ with $L^2(\mathbb R^2)$. 
The goal for the Bayes estimator is to minimize the expected loss over all estimators $\hat x: Y \to X$ that give an estimate of $x$ given measurements $y$. That is,

\begin{equation}
\label{eq:bayesest}
    \hat x_{\text{Bayes}} := \argmin\limits_{\hat x: Y \to X} {\mathbf E}_{x \sim X} (\hat x(y) - x)^2.
\end{equation}

In our approach, we will obtain a neural network approximation to the Bayes estimator $\hat x_{\text{Bayes}}$, where $\argmin$ in Equation \eqref{eq:bayesest} is taken over the estimators given by a family of neural networks and where optimization is performed using minibatch stochastic gradient optimization, with expectation ${\mathbf E}_{x \sim X}$ being approximated by sampling $x \sim X$. For a complete overview on statistical inverse problems, we refer to the book by \cite{Kaipio2005}.

\subsection{DBT neural network}
\begin{figure*}
    \centering
    \includegraphics[trim={0 1.5cm 0 0},clip,width=1.0\textwidth]{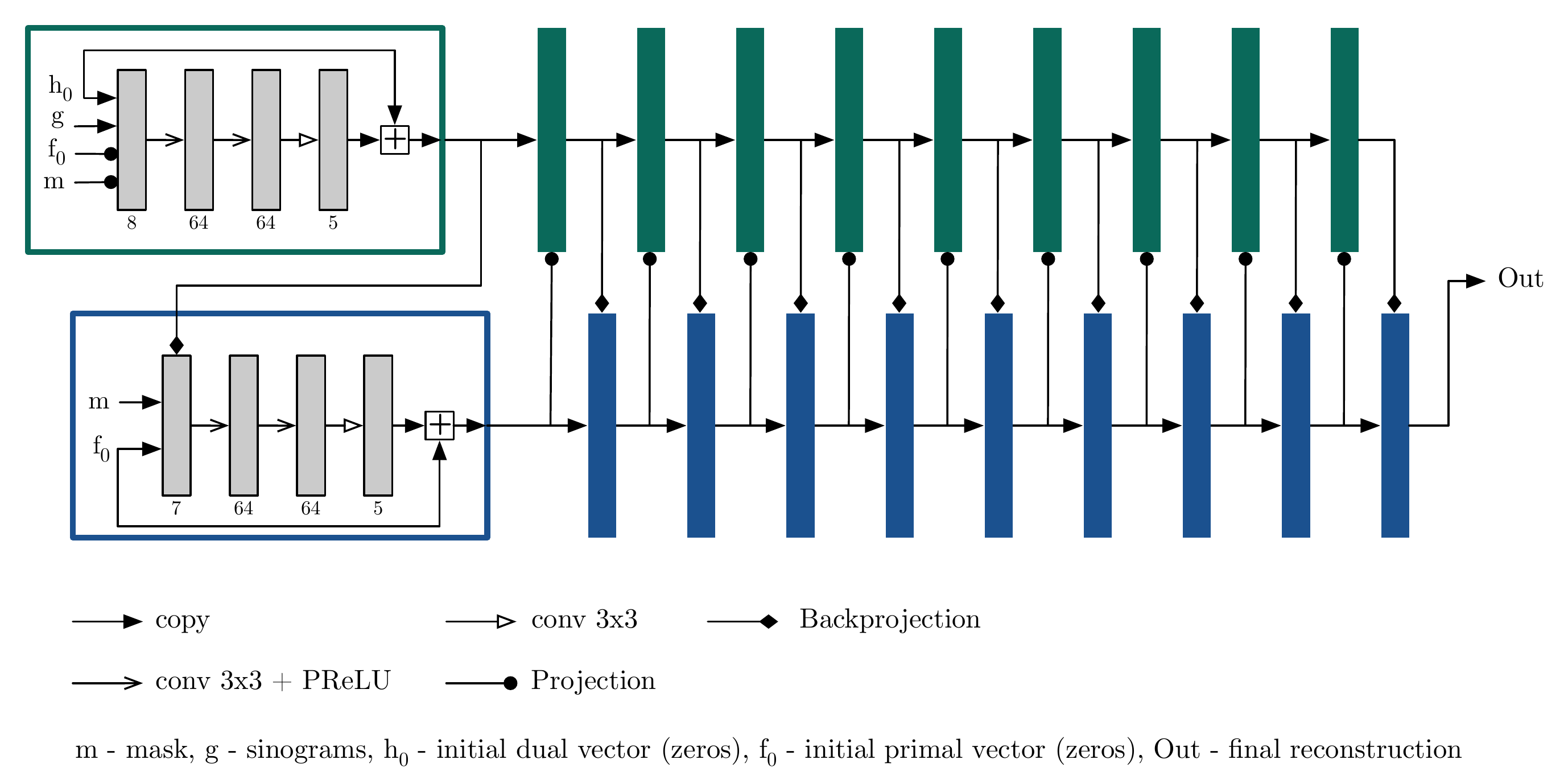}
    \caption{Network architecture of DBToR. Dual blocks (green) are on the upper row and primal blocks (blue) are in the bottom row. The blocks have the same architecture, elaborated in the first blocks. $m$: the breast thickness mask, $h_0$: initial dual vector, $f_0$: initial primal vector, $g$: sinogram data, $\text{Out}$: final reconstruction}
    \label{fig:dbtor}
\end{figure*}
The DBToR algorithm is a data-driven algorithm, which extends the Learned Primal-Dual (LPD) reconstruction algorithm \citep{Adler2017b},  by incorporating additional prior information about the geometry in the form of the thickness measurement of the breast under compression in DBT\footnote{This information is measured by the DBT device and stored in the DICOM image header}. The LPD algorithm itself is an unrolled iterative scheme, based on the proximal primal-dual hybrid gradient method, where proximal operators are replaced by neural networks. The algorithm is trained to reconstruct the images directly from projection data. The DBToR neural network consists of several `reconstruction blocks', which take in projection data, together with information on the thickness of the breast under compression as the initial input, perform a forward and a backward pass by taking projections and back-projections, and use a convolutional neural network to produce an intermediate reconstruction result, which is then improved further by each successive reconstruction block. The architecture and the training algorithm are illustrated in Figure~\ref{fig:dbtor} and Algorithm~\ref{alg:dbt} respectively. In all our experiments we set the number of primal blocks $N_{\textrm{prim}}$ and dual blocks $N_{\textrm{dual}}$ to $10$. The blocks, denoted by $\Gamma_{\theta_i^d}$ and $\Lambda_{\theta_i^p}$ are all ResNet-type blocks consisting of three convolutional layers with kernel size $3 \times 3$ followed by a PreLU layer (with slope initialized as $0.25$) and $64$, $64$ and $5$ filters respectively. As the operators $\mathcal P$ and $\mathcal P^*$ we select the forward and backward operators respectively. The backward operator $\mathcal P^*$ with $\mathcal{P}$ as in \eqref{eq:inv} is the (matrix) adjoint of the forward operator $\mathcal P$. In our model, these are implemented using ASTRA \citep{vanAarle:16}.

At test time, the algorithm takes the input projection data $y$ and the breast thickness information to compute the reconstruction using the function $\texttt{compute\_reconstruction}$.   

In what follows, we denote the training set of objects by $\mathcal D_{\text{train}}$. For an object $x$, we let $y = \texttt{sinogram}_x$ be the corresponding projection data for clarity. As is common, we assume that the training data $\mathcal D_{\text{train}}$ is a representative sample from the domain of DBT images that we want to reconstruct.

The neural network is trained in a supervised fashion as follows. We repeatedly sample an image $x \sim \mathcal D_{\text{train}}$ and the corresponding input projection data $y = \texttt{sinogram}_x$ from the training dataset. The corresponding thickness mask is denoted by $m = \texttt{thickness\_mask}_x$ and is represented by a rectangular mask with the same width as the detector and where the height is given by the measured breast thickness during compression. These measurements are provided by the DBT system, and are available both at training and at test time.

To find the parameters $\theta$ of the neural network $\texttt{compute\_reconstruction}$, we train the network with the $L^2$-loss $l_{\theta} := \| x - z \|_2^2$.  The parameters $\theta$ are updated using the Adam optimizer with a cosine annealing learning rate schedule \citep{Loshchilov2019}, i.e. the learning rate in step $t$ was
\begin{equation*}
	\eta_t = \frac{\eta_0}{2}\Bigl(1 + \cos\Bigl(\pi \frac{t}{N_{\textrm{iter}}} \Bigr) \Bigr)
\end{equation*}
starting at a learning rate $\eta_0$ of $10^{-4}$. For the other Adam parameters, we choose the default parameters of $\beta=(0.9, 0.999)$, $\epsilon =10^{-8}$ and weight decay $0$. The total number of iterations $N_{\textrm{iter}}$ and batch size differ per dataset, and are detailed in the Methods section. Before feeding the data into the network, the input projection data is log-transformed, and scaled such that the standard deviation and mean over the training set is 1.
 

\begin{algorithm}
\centering
\begin{algorithmic}[1]
    \Procedure{\texttt{compute\_reconstruction}}{$m, g$}
        \State $f_0 \gets 0 \in X^{N_{\textrm prim}}$\Comment{Initialize primal vector}
        \State $h_0 \gets 0 \in U^{N_{\textrm dual}}$\Comment{Initialize dual vector}
        \For{$i \gets 0, N$}
        \State $h_i \gets \Gamma_{\theta_i^d} (h_{i-1}, \ \mathcal P(f_{i-1}^{(2)}), \ g,  \ \mathcal P(m)))$
        \State $f_i \gets \Lambda_{\theta_i^p}(f_{i-1}, \ \mathcal P^*(h_i^{(1)}), \ m)$
        \EndFor
        \State \textbf{return} $f_{I}^{(1)}$
    \EndProcedure
    \For{$j \gets 0, N_{\text{iter}} - 1$}
        \State $x \sim \mathcal{D}_{\text{train}}$\Comment{Sample train data}
        \State $y \gets \text{\texttt{sinograms}}_x$\Comment{Sample sinograms}
        \State $m \gets \text{\texttt{thickness\_mask}}_x$\Comment{Create masks}
        \State $z \gets \text{\texttt{compute\_reconstruction}}(m,y)$
        \State $\text{loss} \gets \|z - x\|_2^2$
        \State $\text{change parameters  $\theta_i^p, \theta_i^d, i = 1, \ldots, N$ to reduce loss}$
    \EndFor
\end{algorithmic}
\caption{Pseudocode of the DBToR reconstruction and training algorithm.}
\label{alg:dbt}
\end{algorithm}


  







\subsection{Reconstructed image classification}\label{sec:discretization}
The final step of the reconstruction for estimation of breast density and radiation dose involved in image acquisition, is the classification of the reconstructed image into skin, adipose, and fibroglandular tissue voxels. This is required because rather than relying on voxel attenuation values, which can be quite similar for different types of tissue, we need correct tissue labels to compute the breast density and the dose absorbed by the fibroglandular tissue. Given its high contrast, the skin layer was segmented through a fast seeded region-growing algorithm \citep{Adams1994}, which grows the segmented region starting from a subset of seeds (corresponding to the voxels located on the outer edge of the skin layer) by subsequently including voxels whose intensity was higher than or equal to the mean seed intensity value.
For fibroglandular tissue classification, we extend our previous work on breast CT classification \citep{caballo2018}, and in the first step remove the skin temporarily from the image, with the resulting representation undergoing a well-established, automatic thresholding method based on fuzzy c-means clustering \citep{Bezdek1981}, an algorithm generally adopted in the case of images with low noise content, accompanied by a non-negligible degree of blurring, as is the case for our images. Briefly, voxels are iteratively assigned to a given class (adipose or fibroglandular tissue) in an unsupervised fashion, with the iteration stopping criterion aiming at maximizing the distance between the average voxel values of the two classes. As opposed to traditional cluster analysis, this method allows for a degree of fuzzy overlap between the classes over each iteration, which helps classify the boundary voxels in each subsequent iteration. The fuzzy partition term \citep{Bezdek1981} was experimentally tuned to a value of 2.0.

\section{Materials and Methods}\label{sec:materials}
To train and evaluate the algorithm, we created two datasets of 3D breast phantoms from which we extracted the coronal slices and their corresponding DBT projections. The first dataset consists of virtual 3D breast phantoms generated using a stochastic model, while the second is based on patient dedicated BCT (Breast Computed Tomography) images. DBT projections of these phantoms were simulated using deterministic simulation methods, with the posterior addition of Poisson noise. The use of virtual phantoms not only provided training data, but also allowed for assessing the accuracy of the density and dose estimates, since ground truth is known.

Each voxel of these phantoms was indexed  with a label denoting the corresponding tissue type: skin, adipose tissue, fibroglandular tissue, and Cooper's ligaments. The elemental compositions of these materials were obtained from the work of \cite{Hammerstein1979}, except for the composition of Cooper's ligaments, which was assumed to be identical to that of fibroglandular tissue. 
Linear attenuation coefficients at \SI{20}{\kilo\electronvolt}, a typical average energy of the spectra in clinical DBT imaging, were calculated for each material using the software from Boone and Chavez~\citep{Boone1996}. The resulting linear attenuation coefficients were \SI{0.512}{\per\centi\metre} for adipose tissue, \SI{0.798}{\per\centi\metre} for fibroglandular tissue (and Cooper's ligaments), and \SI{0.854}{\per\centi\metre} for skin.

This process was done for the virtual phantom dataset and the patient BCT dataset.

\subsection{Virtual phantom}
\label{sec:vphantoms}
\begin{figure}[h!]
    \centering
    \includegraphics[width=0.5\textwidth]{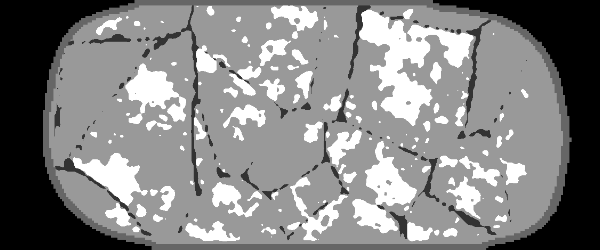}
    \caption{2D coronal breast phantom containing skin (darkest gray), adipose tissue (dark gray), fibroglandular tissue (light gray), and Cooper's ligaments (black).}
    \label{fig:phantom}
\end{figure}

We extracted 41499 coronal slices from 50 breast phantoms generated using the method of \cite{Lau2012}. This method generates breast phantoms in two steps: first, the breast structure is simulated on a coarse scale by generating large compartments of adipose tissue \cite{Zhang2008,Bakic2011}. Second, finer detail for fibroglandular tissue is added subsequently in the form of power-law noise \citep{Reiser2010}. The resulting images have dimensions  $1000 \times 300$ and a resolution of \SI{200}{\micro\metre}. An example of the simulated phantoms is shown in Figure~\ref{fig:phantom}. 
These 41499 coronal slices were used for training and validating the algorithm. Each breast phantom was either included in a training or validation fold completely or not at all, in order to prevent data contamination and bias. 

\subsection{Patient dedicated breast CT phantoms}\label{sec:compression}
Patient dedicated breast CT images were acquired for an unrelated, ethical-board approved patient study evaluating this imaging technology. The images were released for other research purposes after anonymization. In order to compute the density and the accumulated dose to the fibroglandular tissue, the patient breast CT images were automatically classified into four categories (air, skin, adipose and fibroglandular tissue) using a previously developed algorithm for BCT image classification \citep{caballo2018} (see Section~\ref{sec:discretization}). The classified breasts then underwent simulated mechanical deformation as previously described \citep{Fedon2018,Garcia2020}. Briefly, the breasts were converted into a finite element (FE) biomechanical model using the package iso2mesh (v.1.8; Matlab v.13a). A large number of 4-node tetrahedral elements, between 100k and 500k, were used in order to minimize the numerical error during the FE analysis \citep{palomar2008}.
Nearly incompressible (Poisson ratio equal to 0.495), homogeneous and isotropic Neo-Hookean material models for each tissue were used to describe their mechanical behaviour. The Young’s modulus for fibroglandular, adipose and skin tissue were set to \SI{4.46}{\kilo\pascal}, \SI{15.10}{\kilo\pascal}, and \SI{60.00}{\kilo\pascal}, respectively \citep{wellman}. The mechanical compression was then simulated using the open-source package NiftySim (v.2.3.1; University College London) \citep{johnsen2015}, which uses a Total Explicity Dynamic Lagrangian approach to solve the mechanical FE problem \citep{Miller2007}.

Each breast model was compressed to the thickness recorded in the corresponding DICOM header of the cranio-caudal DBT view of that breast, which was acquired, for clinical purposes, during the same visit as the acquisition of the BCT image. 

The total phantom population includes compressed breast thicknesses from \SIrange{3.0}{5.6}{\centi\meter} and chest wall-to-nipple distances from \SIrange{5.8}{18.0}{\centi\meter} with an isotropic voxel size of \SI{0.2}{\milli\metre}$\times$\SI{0.2}{\milli\metre}$\times$\SI{0.2}{\milli\metre}, which is more than sufficient for dosimetric applications \citep{Fedon2018}. 

\begin{figure*}[!t]
    \centering
    \subfloat[]{\includegraphics[width=2.35in]{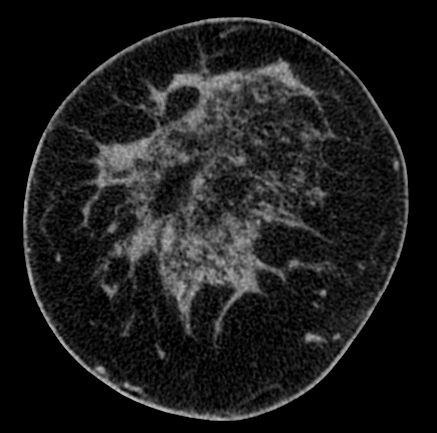}%
    \label{fig:breast-ct-first_case}}
    \hfil
    \subfloat[]{\includegraphics[width=2.35in]{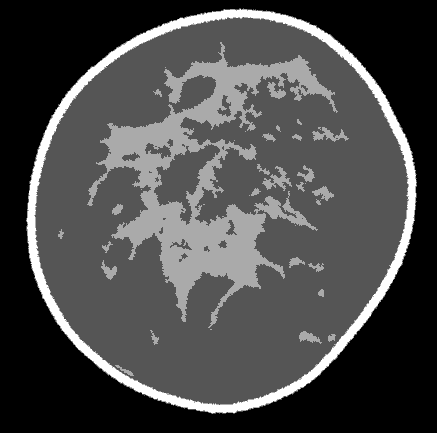}%
    \label{fig:breast-ct-second_case}}
    \hfil
    \subfloat[]{\includegraphics[width=2.35in]{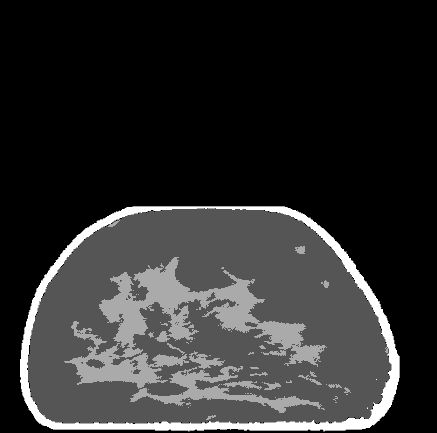}%
    \label{fig:breast-ct-third_case}}
    \caption{(a) Coronal slice of a breast CT image, (b) the same image classified into skin (white), adipose (dark gray) and fibroglandular (light gray) tissue voxels, and (c) the classified deformed image with the technique described in Section~\ref{sec:compression}.}
    \label{fig:breast-ct}
\end{figure*}

Using this method, of which an example is given in Figure~\ref{fig:breast-ct}, we obtained a total of 28891 deformed BCT slices extracted from 91 different patient breasts. Similar to the virtual phantoms the resulting images have dimensions of $1000 \times 300$ pixels and a resolution of \SI{200}{\micro\metre}. Given that the number of deformed BCT slices was substantially lower than that of virtual phantom slices, we pre-trained the model with the latter, and then fine-tuned the model using the BCT slices from 46 patient BCT images. The other 45 patient BCT image phantoms were used for testing the reconstruction performance of the model and the accuracy of density and dosimetry estimates. Each patient breast was either completely included or excluded when selecting slices for fine-tuning and testing the model in order to prevent data contamination and bias.

\subsection{Projection data}
Limited angle fan-beam projections were simulated for all coronal phantom slices using a geometry with the center of rotation placed at the center of the phantom as seen in Figure \ref{fig:mc_geometry}. The x-ray source was placed \SI{65}{\centi\meter} above the center of rotation, and the source-detector distance was \SI{70}{\centi\meter}. A total of 25 equally spaced projections between \SIlist{-24;24}{\degree} were generated, with the detector rotating with the x-ray source. The detector was a perfect photon counting system (100\% efficiency) consisting of 1280 elements with a resolution of \SI{0.2}{\milli\metre}. The forward model \eqref{eq:forward-model} was used for the simulations.

For the virtual phantom data, we generated a series of data sets at 3 noise levels from the noiseless simulated projections. This was accomplished by setting photon count $b_i = 1000 \cdot \sqrt{2}^N$ with $N = \{4, 8, 12\}$. For each noise level, a single Poisson noise realization was generated. For the deformed BCT phantoms, only photon counts corresponding to $N = 8$ were used.

Baseline reconstructions were generated for both noiseless and noisy data using $100$ iterations of the Maximum Likelihood for Transmission (MLTR) algorithm~\citep{Nuyts1998}, using the compressed breast thickness to set the size of the reconstruction volume and with no additional regularization. 

\subsection{Density computation}
Breast density by mass, also called glandularity ($G$), was computed as follows:

\begin{equation}
G =\frac{N_{g} \rho_{g}}{N_{g} \rho_{g} + N_{a} \rho_{a}} 
\end{equation}
where $N_{g}$ and $N_{a}$ are the number of voxels classified as fibroglandular and adipose tissue in the full image, respectively, and $\rho_{g}$ and $\rho_{a}$ are the corresponding density for fibroglandular and adipose tissue, respectively, according to  \cite{Hammerstein1979}. 
The true breast density of each BCT phantom was obtained by applying this equation to the phantom volumes themselves. The estimated breast density resulting from the proposed method was obtained by applying the equation to the classified reconstructed images. 

\subsection{Dose calculation}
\begin{figure}[h!]
    \centering
    \includegraphics[width=0.4\textwidth]{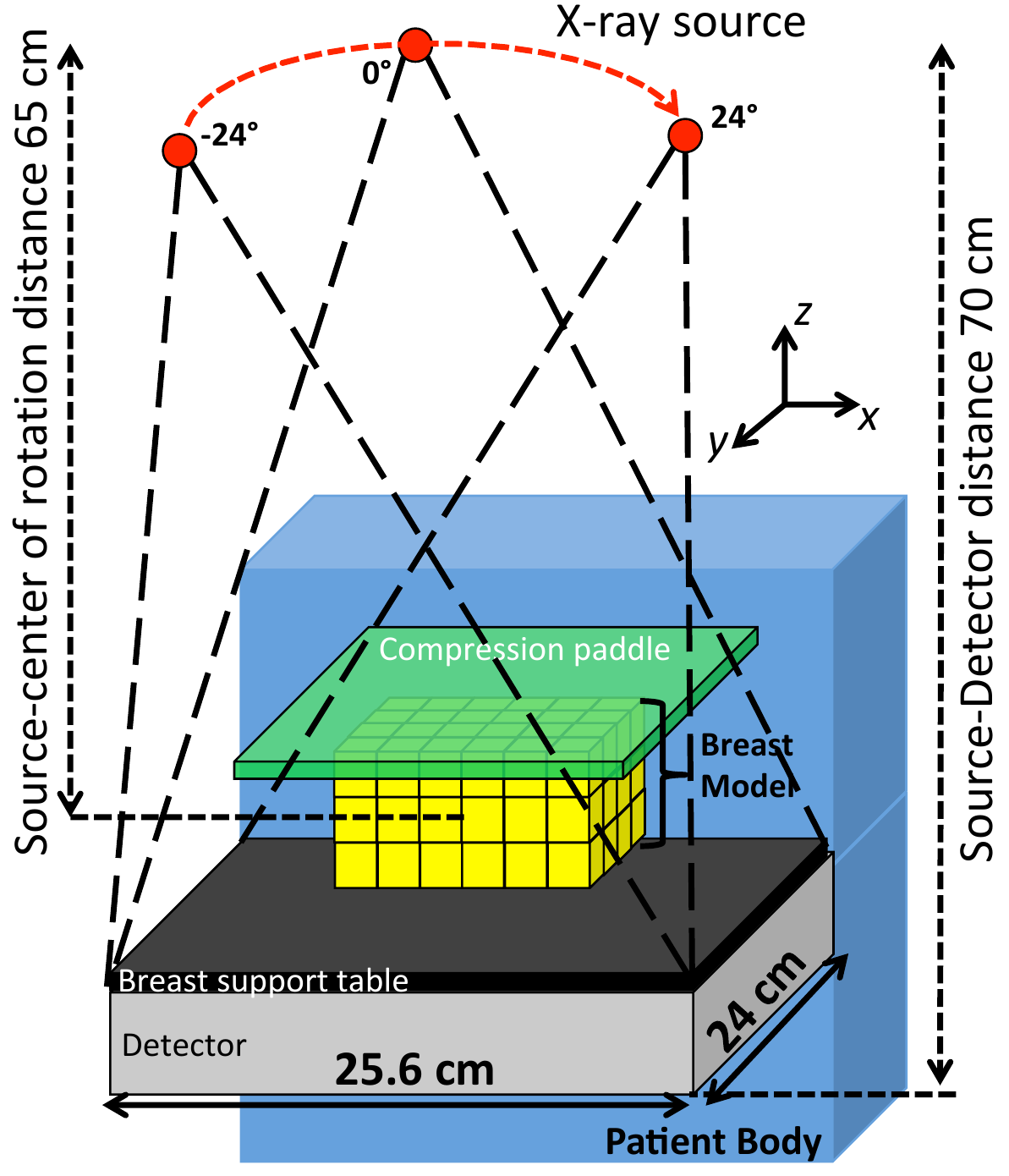}
    \caption{Imaging geometry implemented in the Monte Carlo simulation: the x-ray source is placed at \SI{70}{\centi\meter} from the detector, a \SI{3}{\milli\meter} thick polyethylene terephthalate (PET) compression paddle was simulated and a large water cuboid was included to take into account the patient-body backscatter. The x-ray field irradiated the breast model at different angles (from -24\degree to 24\degree). The center of rotation is placed at 65 cm from the x-ray source. Drawing is not to scale and rotation of the detector is not shown.}
    \label{fig:mc_geometry}
\end{figure}

The mean glandular dose ($\text{MGD}$) estimations were performed using a previously described and validated Monte Carlo code \citep{fedon2018_homo,fedon2018_hetero}, based on the Geant4 toolkit (release 10.05, December 2018).
The Monte Carlo geometry replicates the one used to generate the projections and is shown in Figure~\ref{fig:mc_geometry}. 

As during simulation of the DBT projections above, each voxel was labelled with an index related to its composition: air, adipose tissue, fibroglandular tissue, and skin, using the chemical compositions reported by \cite{Hammerstein1979}.
The energy deposited in the fibroglandular voxels was recorded and then converted into dose according to the formula
\begin{equation}
\text{MGD} =\frac{\sum_i E_{i}}{M_{g}} 
\end{equation}
where $E_{i}$ is the energy deposited at the interaction event $i$, and $M_{g}$ is the total fibroglandular breast mass.

\begin{table}[t!]
\caption{X-ray spectra used in the Monte Carlo simulation. HVL: 1st half value layer}
\label{tab_spectrum}
\centering
\begin{tabular}{llll}
\toprule
Breast  & Spectrum & HVL  & \# Cases \\
Thickness (mm) & & (mm Al) &  \\
\midrule
30-39 & W/Rh - 27 kV  & $0.519$ & $2$\\
40-49 & W/Rh - 28 kV  & $0.530$ & $7$\\
50-59 & W/Rh - 29 kV  & $0.538$ & $6$\\
60-69 & W/Rh - 30 kV  & $0.547$ & $18$\\
70-79 & W/Rh - 31 kV  & $0.557$ & $12$\\
\bottomrule
\end{tabular}
\end{table}

$10^7$ primary x-rays were emitted by an isotropic point source placed at \SI{70}{\centi\metre} from the detector and collimated to irradiate the entire detector. In order to replicate the tomosynthesis acquisition mode, a total of 25 projections were simulated from $-24\degree$ to $24\degree$, every 2\degree. The projection at $0\degree$ replicates the mammographic acquisition. The number of primary particles ensured a statistical uncertainty on the total dose of less than $0.7\%$, evaluated using the method proposed by \cite{sempau2001}.
Photoelectric interactions, and coherent and incoherent scatter were included in the simulations without modifying the default cut range for photons (\SI{1}{\milli\metre}, corresponding to an energy of \SI{2.45}{keV} and \SI{2.88}{keV} for adipose and fibroglandular tissue, respectively).

The x-ray spectra were modeled using the TASMICS model \citep{hernandez2014} by adjusting the thickness of the modeled rhodium filter to match the first half-value layer measured with a solid state detector (RaySafe X2-MAM sensor, Billdal, Sweden) in the modelled system as shown in Table \ref{tab_spectrum}. 
The Monte Carlo simulations for estimating the MGD were performed twice for each breast; once for each of the 45 BCT phantoms, and once each for the corresponding labelled DBToR reconstructions. In this way, the accuracy of the resulting patient-specific dosimetry estimates could be assessed.  

\subsection{Comparison to current reconstructions}
The results of the DBToR reconstruction, prior to the voxel classification for estimation of breast density and dose, were compared to the baseline iterative MLTR reconstruction algorithm, the LPD algorithm and the U-Net trained on FBP reconstructions for both noiseless and noisy data.

\subsection{Model training and evaluation}
In total, we trained three versions of the DBToR algorithm: two versions were trained on virtual phantom  projections, both without any noise and with varying levels of noise, and one version was pretrained on noisy virtual phantom projections and subsequently finetuned on noisy deformed BCT phantoms, on the data described in Sections \ref{sec:vphantoms} and \ref{sec:compression}. For comparison, we trained the basic LPD algorithm (achieved by removing the height mask from the input) in addition to DBToR on virtual phantom projections. Finally, we also trained U-Net baselines on FBP reconstructions of DBT slices as a classical deep learning reconstruction baseline. In our U-Net baselines, the standard U-Net architecture \cite{Ronneberger2015a} with depth $4$, $32$ filters in the initial double convolution block and instance normalization layers was used.

For the DBToR and LPD models, we used a batch size of $8$ for the pretraining on the virtual phantoms, and for the finetuning on the BCT phantoms for the dosimetry application, we chose a batch size of $1$. For the U-net baseline, we use a batch size of $4$. The number of iterations $N_{\text{iter}}$ was set to $10^5$ for all models trained on virtual phantoms, and to $4 \cdot 10^5$ for the final model finetuning on BCT phantoms. This leads to a training time to about 48 hours and 24 hours for the DBToR/LPD and the U-net models respectively. At inference stage, the DBToR reconstruction takes less than $1$ second for a single coronal slice. 

We use the $L^2$ loss, Structural Similarity Index (SSIM) \citep{Wang2004} and Peak Signal-to-Noise Ratio (PSNR) as performance metrics.

\section{Results}\label{sec:results}
For DBToR, LPD and U-Net trained on noise-free virtual phantom data we report the corresponding $L^2$ loss, SSIM and PSNR on noise-free virtual phantom test data in Table \ref{tab:comparison-between-algos}. Lower $L^2$ loss and higher SSIM and PSNR values indicate better reconstruction performance. For DBToR, LPD and the U-Net trained on noisy virtual phantom projections, we report these metrics for noise levels $N=4, 8, 12$ in Table \ref{tab:comparison-between-algos-noisy}. In both tables, we report the mean and standard deviation of the metrics obtained using $3$ cross-validation folds with $60\%$ of the data used for training and $40\%$ used for testing in each fold.

The original LPD algorithm is significantly outperformed by DBToR for all noise levels. Visual inspection of the slices produced by LPD revealed that the LPD often reconstructs breast regions adjacent to the compression paddle very poorly for both test and training data. In particular, it frequently fails to reproduce the flatness of the part of the skin surface which is in contact with the compression paddle. We ruled out overfitting of LPD as the cause of the artifacts, since the validation metrics remained low throughout the training process. While it is possible that a much larger version of LPD with more reconstruction blocks would learn to correct these artifacts, we will see that DBToR resolves them without any further increase in the number of parameters. 
The U-Net architecture is outperformed by DBToR for noise levels $N=4$ and $N=8$, but it outperforms DBToR in terms of PSNR for noise level $N=12$ and in the noise-free case. However, U-Net yields reconstructions that suffer from noticeably lower SSIM compared to DBToR, and a visual inspection of the volumes revealed that the U-Net also struggles with reconstructing the breast shape correctly, which can have a strong negative effect on the dosimetry accuracy. Higher PSNR for the U-Net can be, potentially, explained by the fact that U-Net reconstructs the background better, which is not important for dose computations. Additionally, the variance in the performance metrics is higher for the U-Net, indicating that it is more sensitive to the data split. Overfitting was more pronounced for the U-Net baseline due to the much higher parameter count (7.8M parameters for U-Net and 872k parameters for DBToR). These considerations make DBToR a preferred candidate for our application.
The proposed DBToR algorithm also outperforms the iterative MLTR reconstruction algorithm at all noise levels and for all metrics being considered, while yielding visually more accurate reconstructions as well. It is also interesting to note from Table \ref{tab:comparison-between-algos-noisy} that the performance of DBToR at noise level $N=4$ is superior to the MLTR reconstruction algorithm at noise level $N=12$, which corresponds to 8 times higher photon count. At noise level $N=12$, the performance of MLTR is slightly below the performance of MLTR for the noise-free case. At the same time, DBToR at $N=12$ reaches comparable level of performance to that of the DBToR on noise-free data.

\begin{figure}[h!]
    \centering
    \includegraphics[trim={8cm 0.75cm 8cm 0},width=0.33\textwidth,clip]{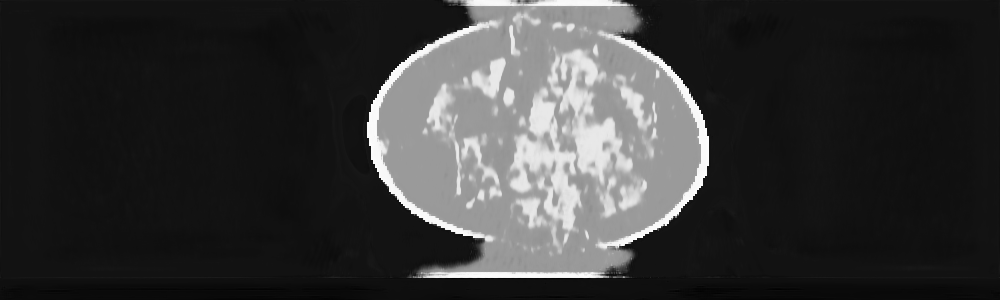}
    \caption{Example of U-Net reconstruction artifact}
    \label{fig:unetsample}
\end{figure}

\begin{table*}[t!]
\caption{Results on noise-free phantom projections, mean $\pm$ standard deviation (in bold best result)}
\label{tab:comparison-between-algos}
\centering
\begin{tabular}{>{}l>{}l>{}l>{}l}
\toprule
Model & $L^2$-loss & SSIM & PSNR \\
\midrule
MLTR &  $0.007 \pm 4.3 \cdot 10^{-5}$ & $0.83 \pm 4.2 \cdot 10^{-3}$ & $20.2 \pm 2.9 \cdot 10^{-2}$ \\
LPD & $0.014 \pm 6.9 \cdot 10^{-3}$ & $0.87 \pm 2.9 \cdot 10^{-2}$ & $19.3 \pm 1.8$ \\
U-Net & $\mathbf{0.002 \pm 7.0 \cdot 10^{-4}}$ & $0.80 \pm 1.4 \cdot 10^{-1}$ & $\mathbf{27.8 \pm 2.3}$ \\
DBToR & $0.003 \pm 8.1 \cdot 10^{-4}$ & $\mathbf{0.91 \pm 2.3 \cdot 10^{-2}}$ & $24.8 \pm 1.3$\\
\bottomrule
\end{tabular}
\end{table*}

\begin{table*}[t!]
\caption{Results on noisy phantom projections for different noise levels $N$, mean $\pm$ standard deviation (in bold best result)}
\label{tab:comparison-between-algos-noisy}
\centering
\begin{tabular}{>{}l>{}l>{}l>{}l}
\toprule
Model & $L^2$-loss & SSIM & PSNR \\
\midrule
MLTR ($N=4$) & $0.0096 \pm 1.0 \cdot 10^{-4}$ & $0.69 \pm 9.0 \cdot 10^{-3}$ & $19.17 \pm 5.8 \cdot 10^{-2}$ \\
LPD ($N=4$) & $0.014 \pm 5.4 \cdot 10^{-3}$ & $0.85 \pm 2.4 \cdot 10^{-2}$ & $18.79 \pm 1.6$ \\
U-Net ($N=4$) & $0.0085 \pm 1.4 \cdot 10^{-3}$ & $0.64 \pm 1.7 \cdot 10^{-1}$ & $20.96 \pm 1.0$ \\
DBToR ($N=4$) & $\mathbf{0.0044 \pm 9.1 \cdot 10^{-4}}$ & $\mathbf{0.90 \pm 1.8 \cdot 10^{-2}}$ & $\mathbf{23.39 \pm 9.6 \cdot 10^{-1}}$ \\
\midrule
MLTR ($N=8$) & $0.0082 \pm 5.2 \cdot 10^{-5}$ & $0.74 \pm 7.2 \cdot 10^{-3}$ & $19.87 \pm 3.6 \cdot 10^{-2}$ \\
LPD ($N=8$) & $0.013 \pm 5.3 \cdot 10^{-3}$ & $0.84 \pm 3.0 \cdot 10^{-2}$ & $19.06 \pm 1.6$ \\
U-Net ($N=8$) & $0.0043 \pm 1.6 \cdot 10^{-3}$ & $0.78 \pm 1.3 \cdot 10^{-1}$ & $24.13 \pm 1.45$ \\
DBToR ($N=8$) & $\mathbf{0.0033 \pm 6.3 \cdot 10^{-4}}$ & $\mathbf{0.91 \pm 1.2 \cdot 10^{-2}}$ & $\mathbf{24.71 \pm 8.0 \cdot 10^{-1}}$ \\
\midrule
MLTR ($N=12$) & $0.0078 \pm 4.47 \cdot 10^{-5}$ & $0.80 \pm 5.5 \cdot 10^{-3}$ & $20.06 \pm 3.0 \cdot 10^{-2}$ \\
LPD ($N=12$) & $0.015 \pm 6.3 \cdot 10^{-3}$ & $0.86 \pm 2.7 \cdot 10^{-2}$ & $19.04 \pm 1.6$ \\
U-Net ($N=12$) & $0.0034 \pm 1.2 \cdot 10^{-3}$ & $0.72 \pm 1.6 \cdot 10^{-1}$ & $\mathbf{25.58 \pm 2.3}$ \\
DBToR ($N=12$) & $\mathbf{0.0034 \pm 9.4 \cdot 10^{-4}}$ & $\mathbf{0.91 \pm 1.6 \cdot 10^{-2}}$ & $24.47 \pm 1.6$ \\
\bottomrule
\end{tabular}
\end{table*}

\begin{table}[t!]
\caption{Results on noisy BCT phantom projections (in bold best result)}
\label{tab:comparison-between-algos-bct}
\centering
\begin{tabular}{llll}
\toprule
Model & $L^2$-loss & SSIM & PSNR \\
\midrule
MLTR & $0.006$ & $0.82$ & $21.45$\\
DBToR & $\mathbf{0.0018}$ & $\mathbf{0.93}$ & $\mathbf{27.03}$\\
\bottomrule
\end{tabular}
\end{table}

\begin{figure*}[!ht]
    \centering
    \subfloat[]{\includegraphics[width=0.40\linewidth]{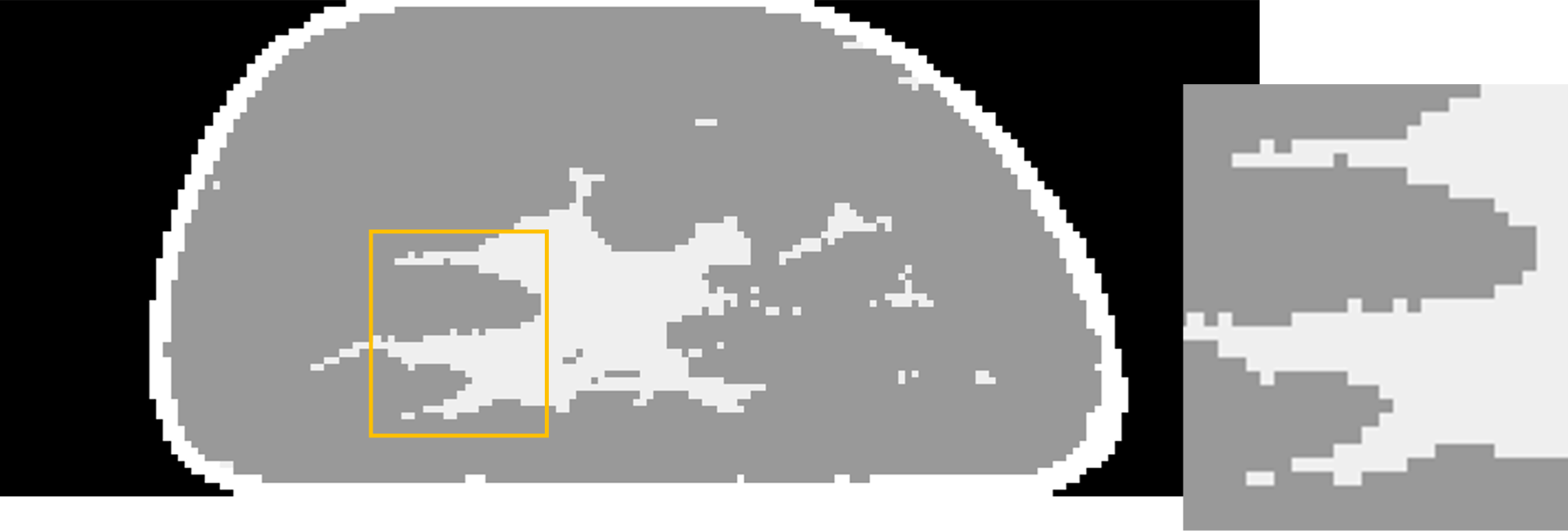}%
    }
    \hfil
    \subfloat[]{\includegraphics[ width=0.40\linewidth]{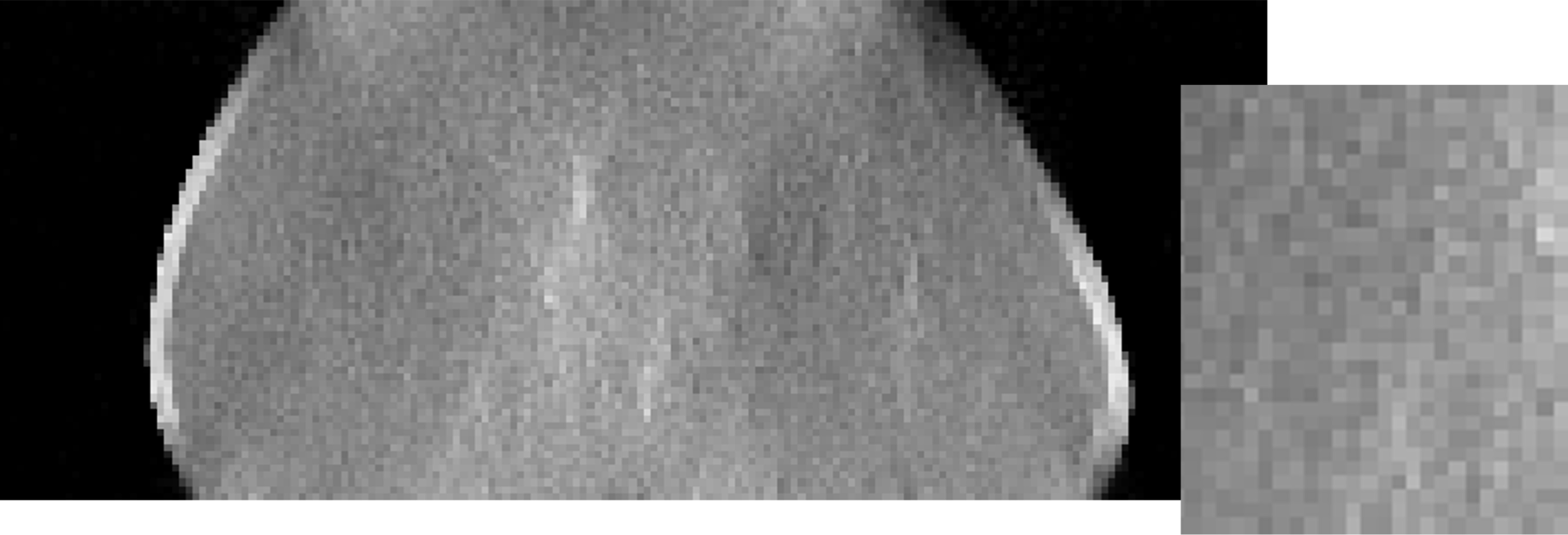}%
    }
    \hfil
    \subfloat[]{\includegraphics[ width=0.40\linewidth, ]{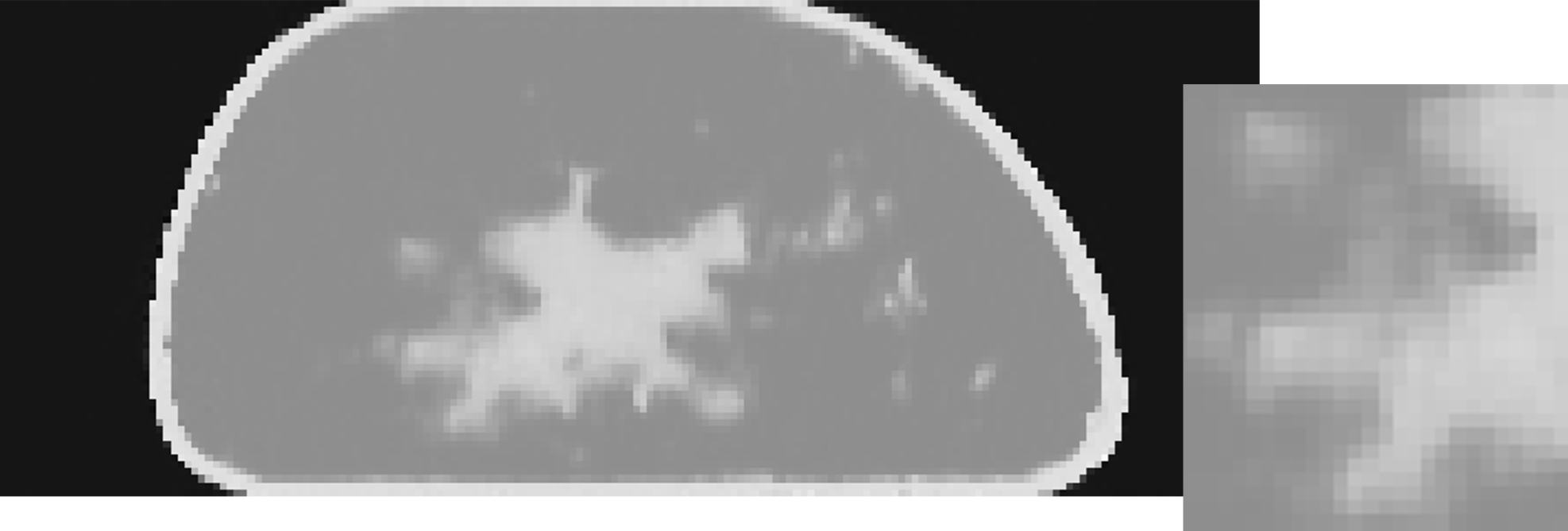}%
    }
    \hfil
    \subfloat[]{\includegraphics[ width=0.40\linewidth]{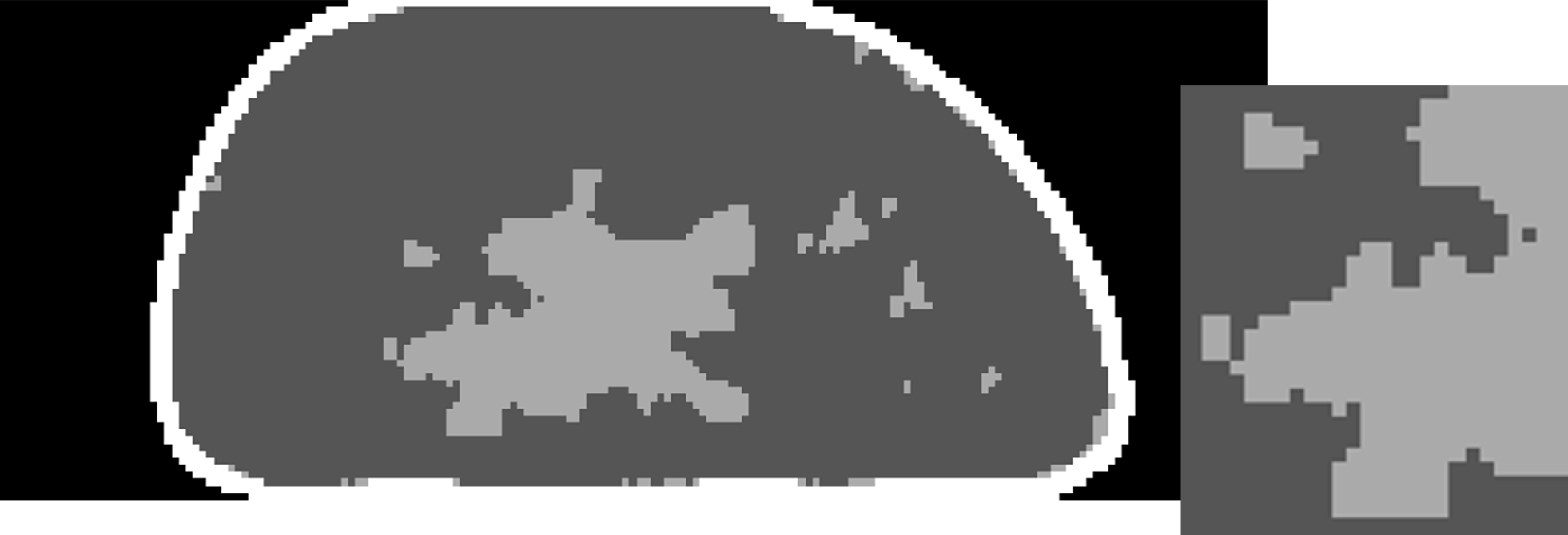}%
    }
    \caption{(a) Coronal slice of a breast CT phantom, (b) MLTR reconstruction, (c) DBToR reconstruction, and (d) classification of DBToR reconstruction.}
    \label{fig:dbt-s1}
\end{figure*}

\begin{figure}[!ht]
    \centering
    \subfloat[]{\includegraphics[width=0.5\linewidth]{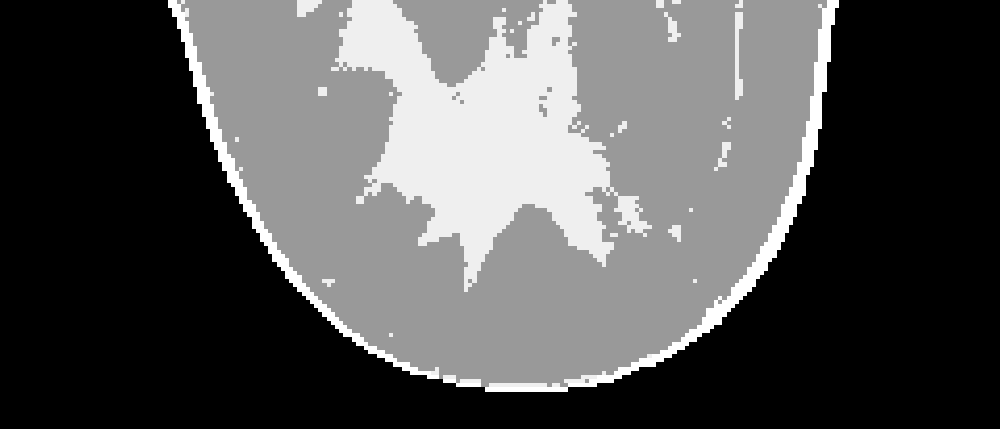}%
    }
    \hskip -4ex
    \subfloat[]{\includegraphics[width=0.5\linewidth]{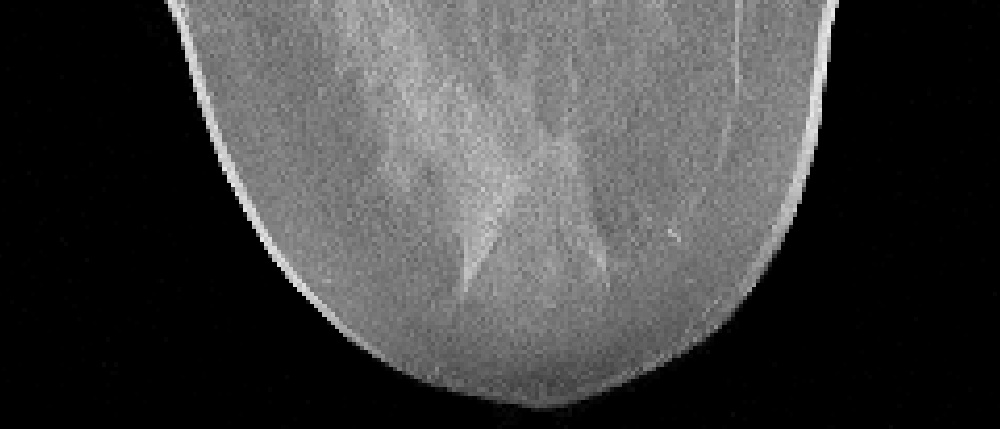}%
    }
    \hfil
    \subfloat[]{\includegraphics[width=0.5\linewidth]{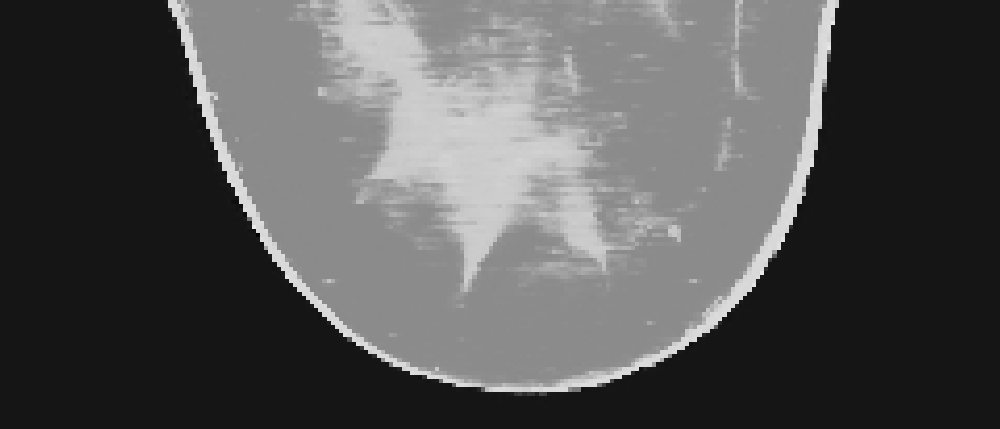}%
    }
    \hskip -4ex
    \subfloat[]{\includegraphics[width=0.5\linewidth]{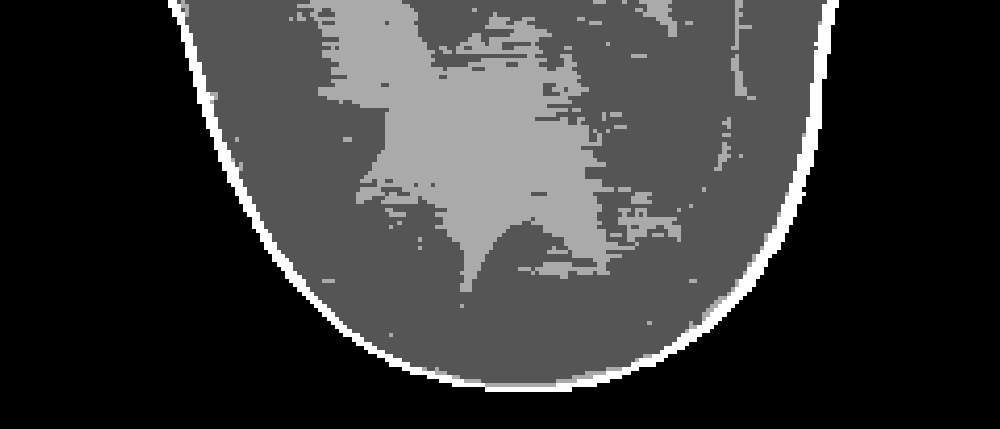}%
    }
    \caption{(a) Axial slice of a breast CT phantom, (b) MLTR reconstruction, (c) DBToR reconstruction, and (d) classification of DBToR reconstruction. This view is created from the volume assembled by stacking all coronal slices from the same case.}
    \label{fig:dbt-s2}
\end{figure}

\begin{figure*}[!ht]
    \centering
    \subfloat[]{\includegraphics[width=0.245\linewidth]{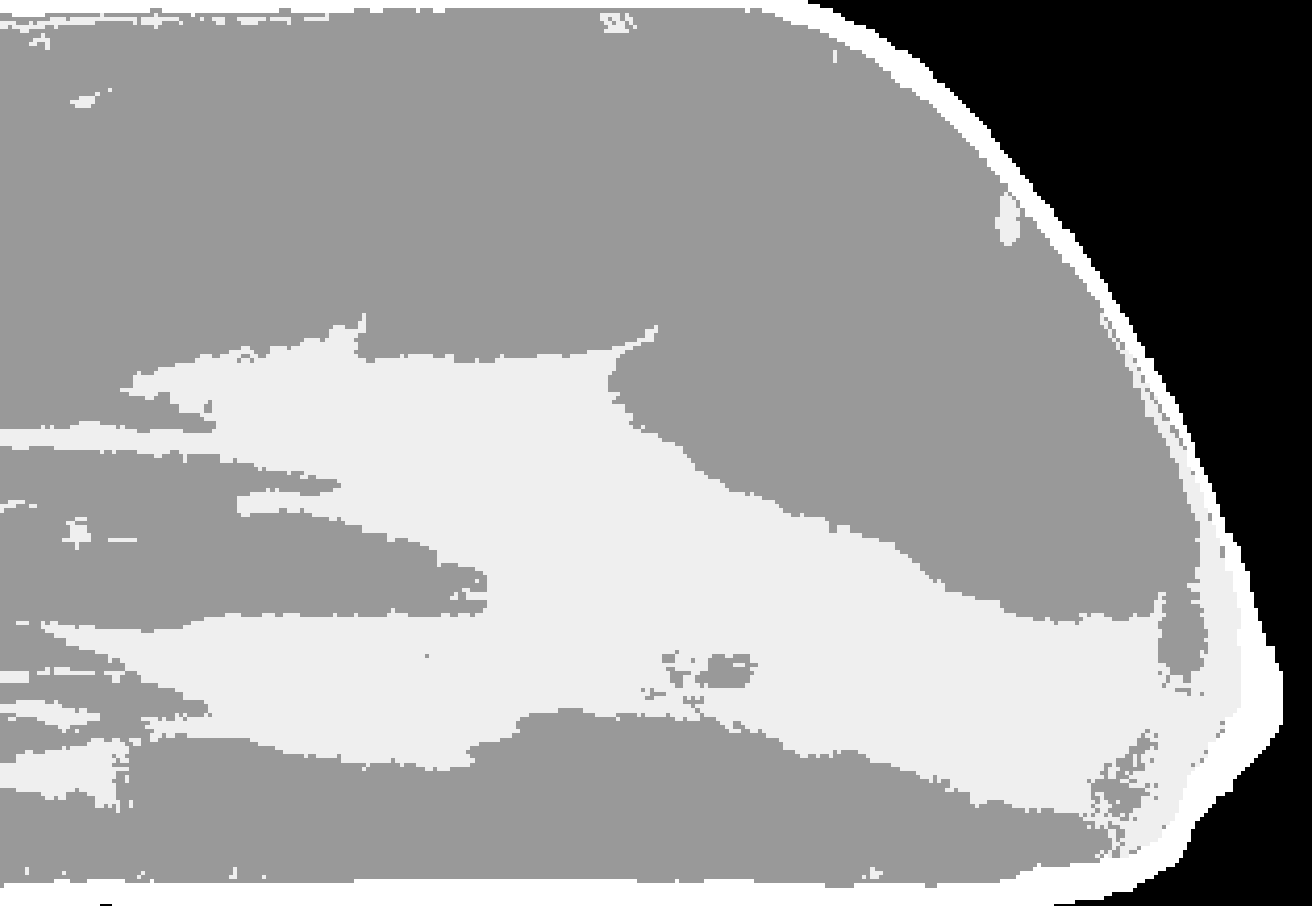}%
    }
    \hfil
    \subfloat[]{\includegraphics[width=0.245\linewidth]{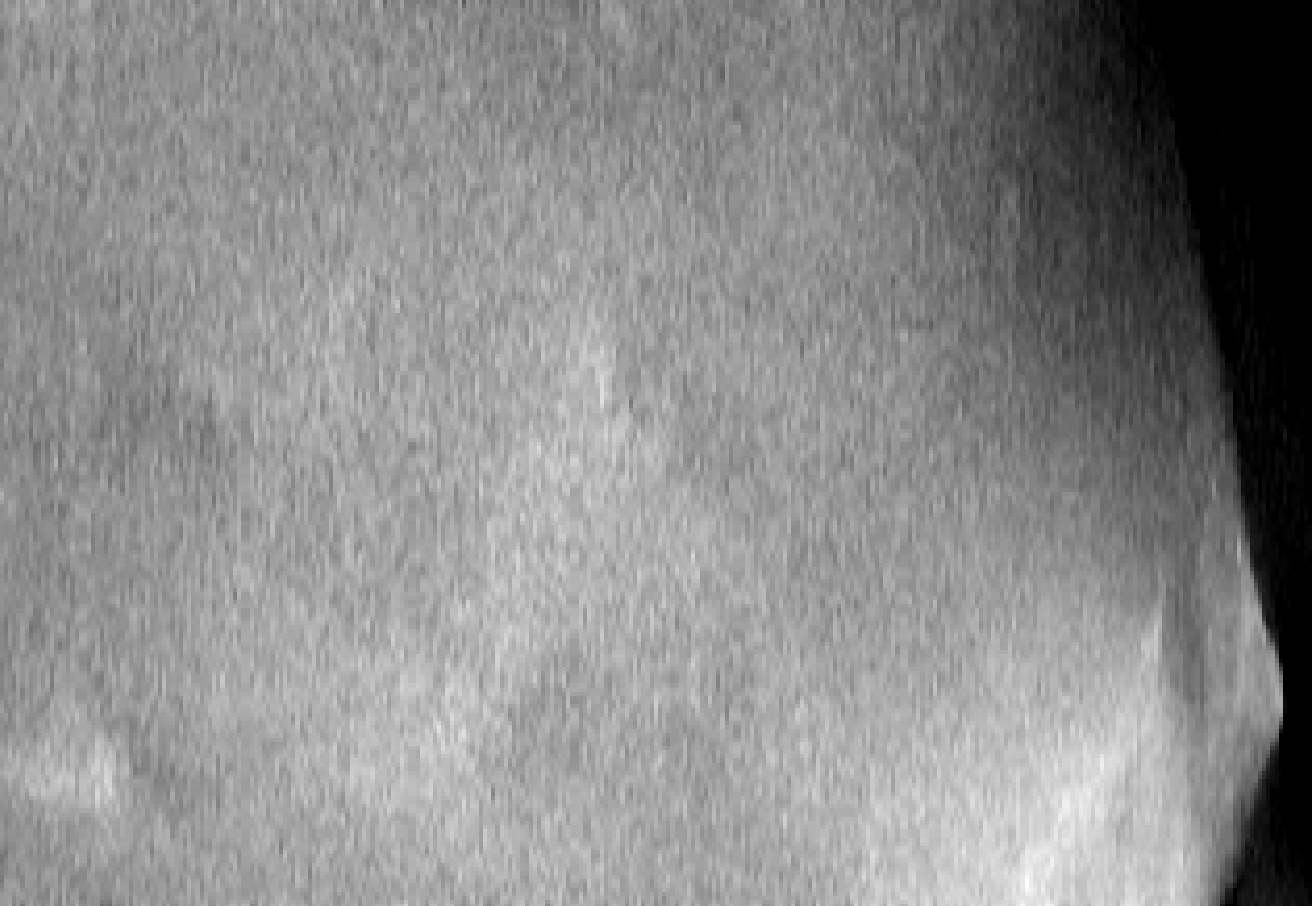}%
    }
    \hfil
    \subfloat[]{\includegraphics[width=0.245\linewidth]{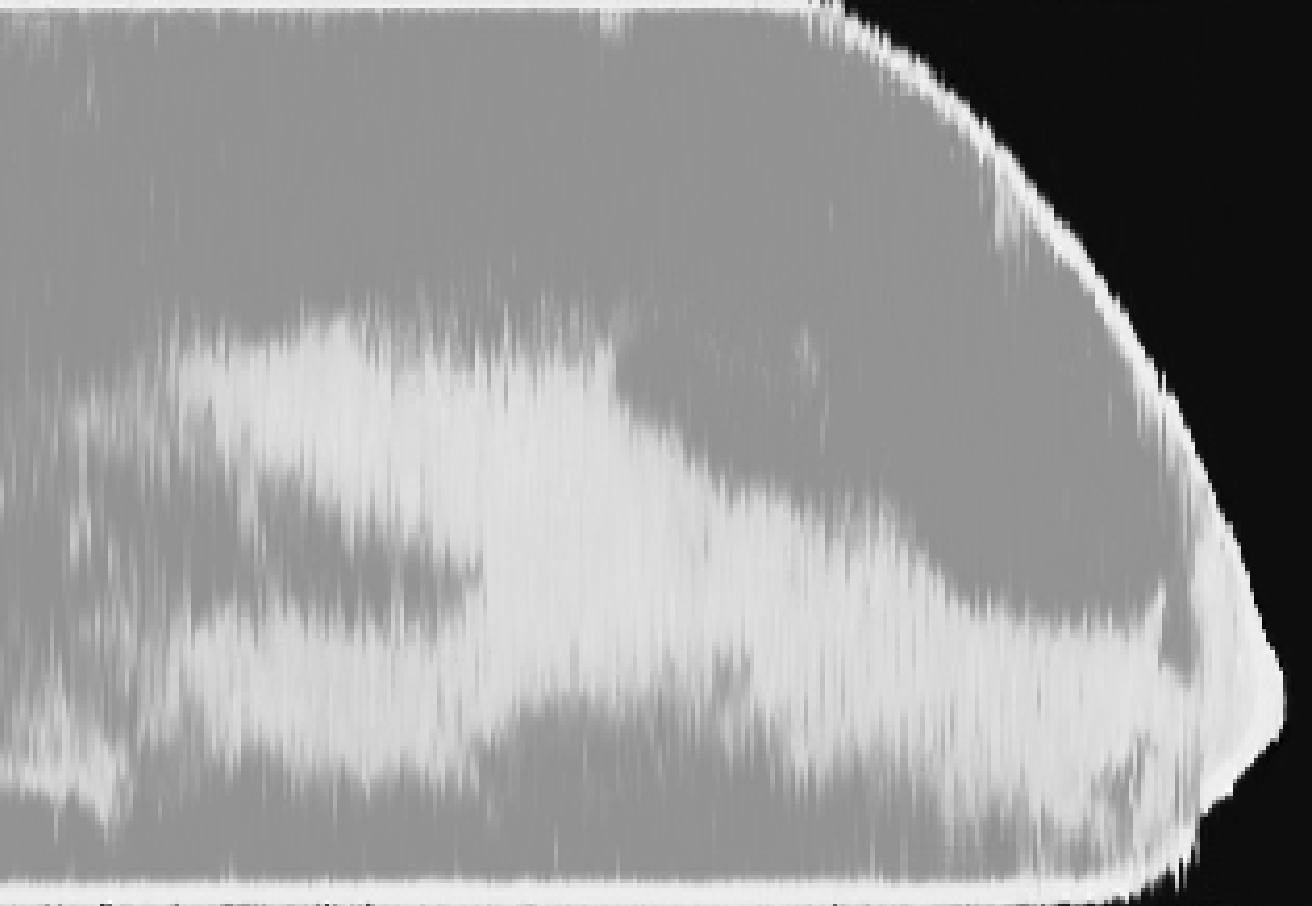}%
    }
    \hfil
    \subfloat[]{\includegraphics[width=0.245\linewidth]{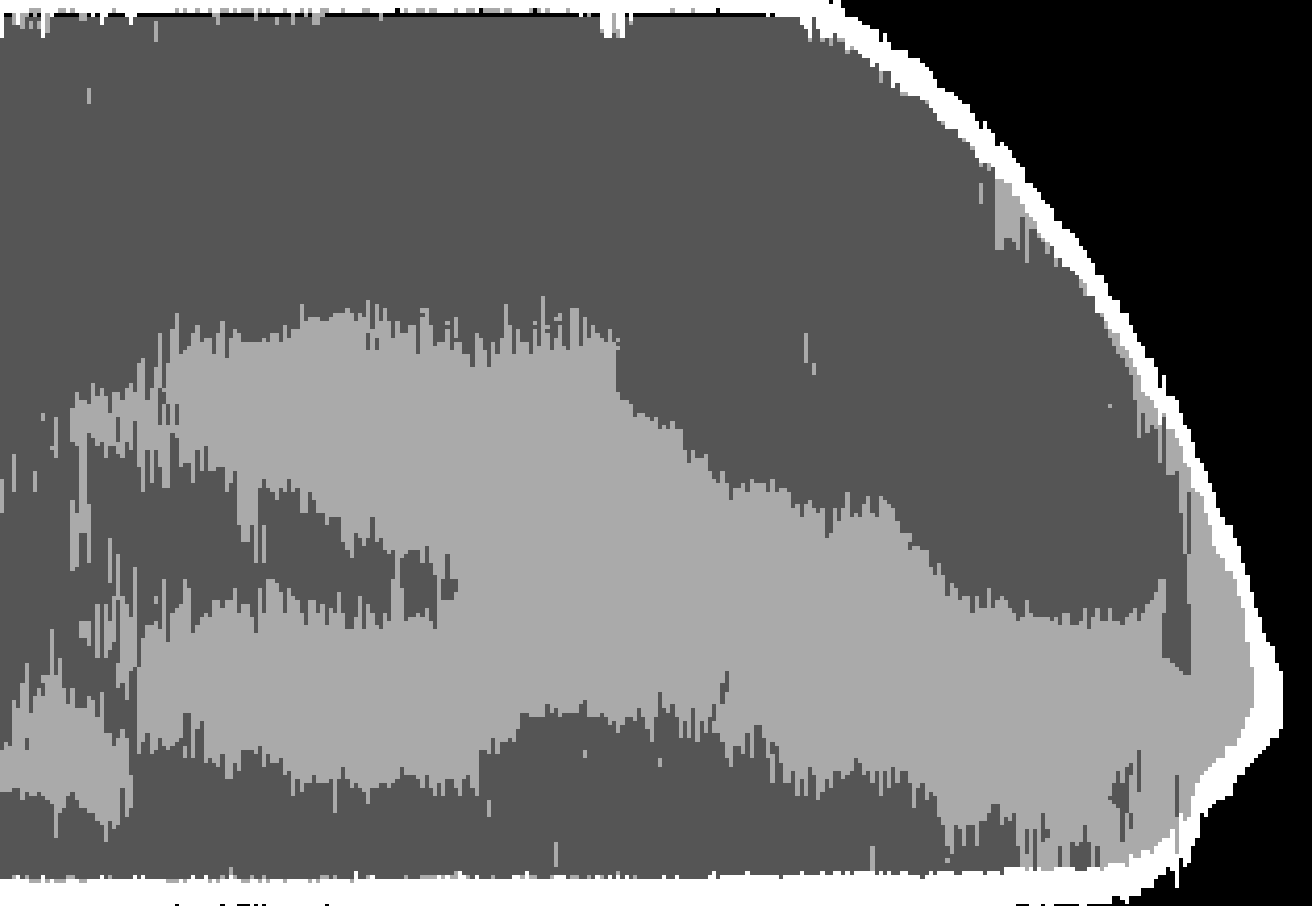}%
    }
    \caption{(a) Sagittal slice of a breast CT phantom, (b) MLTR reconstruction, (c) DBToR reconstruction, and (d) classification of DBToR reconstruction. This view is created from the volume assembled by stacking all coronal slices of this case.}
    \label{fig:dbt-s3}
\end{figure*}

For DBToR trained on noisy virtual phantom projections and subsequently finetuned on deformed breast CT slices, where we used noise level $N=8$ only during training and finetuning, we summarize the reconstruction performance in Table \ref{tab:comparison-between-algos-bct}, and in Figures \ref{fig:dbt-s1}, \ref{fig:dbt-s2}, and \ref{fig:dbt-s3} we give examples of coronal, axial, and sagittal slices of the virtual breast phantom and corresponding MLTR reconstruction, DBToR reconstruction, and DBToR classification (all on noisy deformed breast CT slices). We observe that DBToR outperforms the baseline MLTR algorithm in terms of the reported metrics and visual quality of the slices, particularly noticeable for coronal and sagittal directions.

\subsection{Breast density estimation}
\begin{figure}
\centering
   \includegraphics[width=1\linewidth]{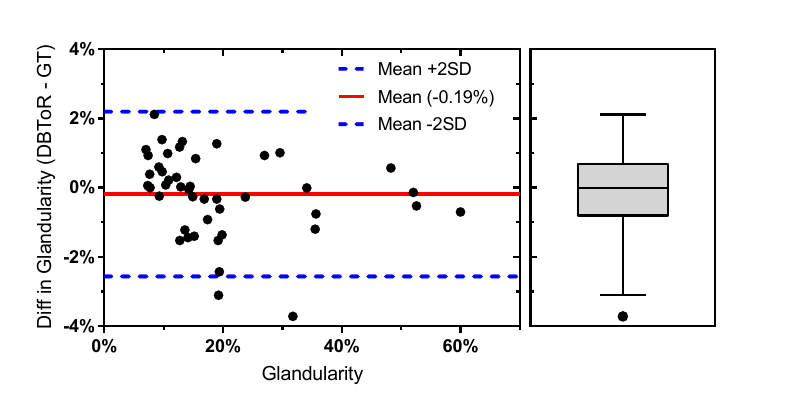}
   \caption{Bland-Altman plot of the difference (DBToR - GT) of glandularity (in percentage points) and matching box plot with Tukey whiskers.}
   \label{fig:diff_gland}
\end{figure}

Figure~\ref{fig:diff_gland} shows a box-whisker plot of the absolute percentage difference in glandularity between the DBToR estimate and the ground truth (GT). It can be seen that, on average, no bias is observed ($p$-value $0.3$), and that the breast density estimates are accurate to within 2\%.


\subsection{Mean glandular dose estimation}
\begin{figure*}[!t]
    \centering
    \subfloat[]{\includegraphics[width=3.5in]{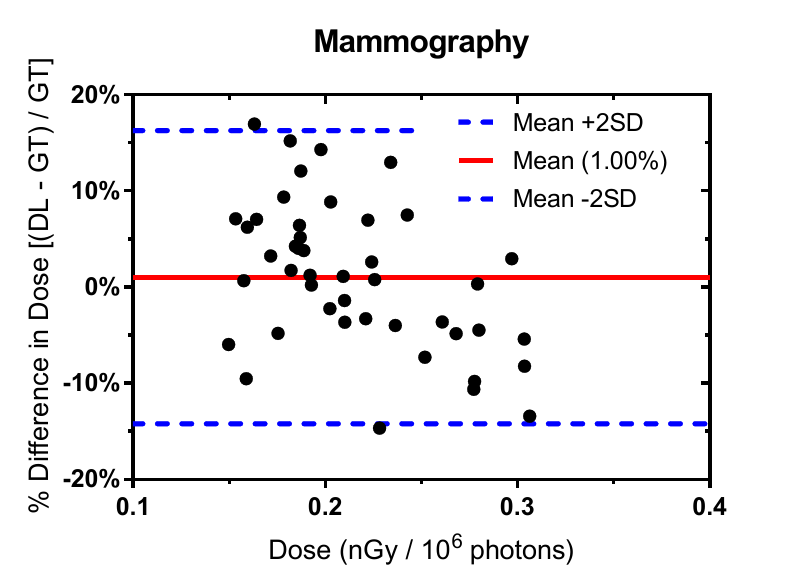}}%
    \hfil
    \subfloat[]{\includegraphics[width=3.5in]{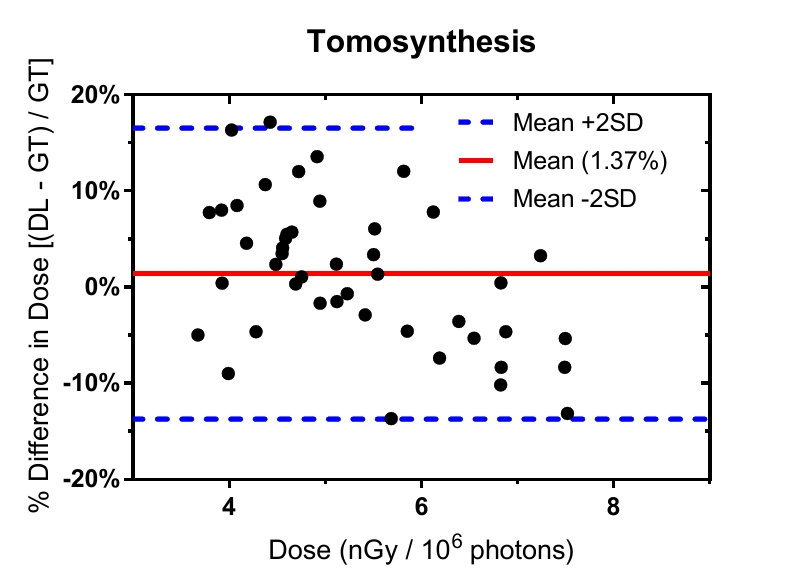}}%
    \caption{Bland-Altman plot of the difference in dose estimates resulting from the DBToR reconstruction and the ground truth, in percentage, for both mammography (a) and DBT (b). The red line represents the mean, while the two blue-dashed lines represent the 95\% limits of agreement.}
    \label{fig:BA_plot}
\end{figure*}

The comparison between the MGD evaluated from the DBToR reconstructions and the GT is shown in the Bland-Altman plots of Figure~\ref{fig:BA_plot}, for both mammography and DBT geometries.

\begin{figure}[h!]
    \centering
    \includegraphics[width=0.5\textwidth]{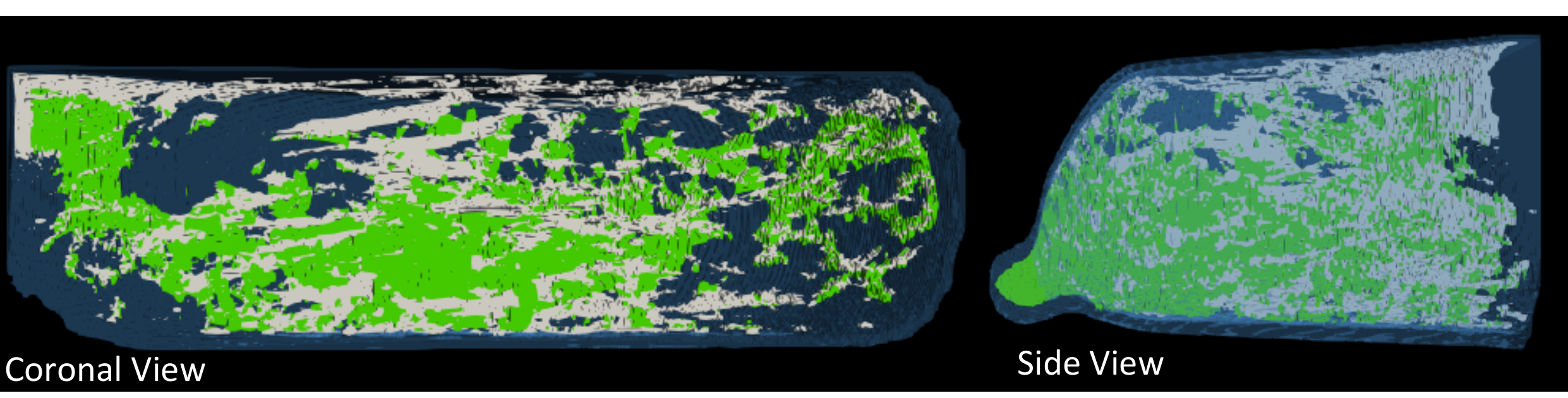}
    \caption{Example of fibroglandular tissue distribution for an outlier case: the ground truth (in white) depicts a higher amount of fibroglandular tissue in the top breast layer (i.e., facing the x-ray tube), while the DBToR model (in green) predicts a fibroglandular distribution spread towards the anterior, and more inferior, part of the breast.}
    \label{fig:diff_distribution}
\end{figure}

As can be seen, no bias is observed in the proposed dose estimation method ($p$-value $0.23$), with the data points equally scattered around zero, and that the largest error in the dose estimation is less than 20\%.
Visual inspection of the cases that lie beyond the $\pm$2SD limits reveals that this is the consequence of differences in the reconstructed fibroglandular distribution obtained with the DBToR model compared to the GT, as shown in an example in Figure~\ref{fig:diff_distribution}.

For this case, the absolute difference on the glandularity is $1.5\%$ (namely $12.0\%$ for the DBToR and $13.5\%$ for GT). Thus, due to a higher amount of fibroglandular tissue closest to the x-ray source in the case of the GT, the radiation dose estimated by DBToR is lower by $13.5\%$.

\section{Discussion and conclusion}\label{sec:conclusion}
We presented a deep learning-based method for the reconstruction of DBT, which we call DBToR. The model is both data driven and model-based, since the forward and backprojection operators for a given DBT geometry are a part of our neural network architecture and at the same time the model is trained to reduce the tomographic artifacts of the reconstruction. As training data we used two sources of data, one based on random samples with statistical properties similar to real breast volumes, and one dataset of patient breast CT images that have been compressed with a finite element model to simulate the same breast under compression in a DBT system. 

Compared to LPD, in DBToR we added the compressed breast thickness as prior information. Since the limited angle causes a severely ill-posed problem, it was expected that this information would be definitely required, and the experiments (Table \ref{tab:comparison-between-algos}) confirmed that the result dramatically degrades, compared to the 'full' DBToR, when this prior knowledge is not provided to the algorithm. Requiring this additional information does not limit the generalizability of the method, since it is readily available in all DBT systems.

The results indicate that the proposed algorithm outperforms the MLTR iterative reconstruction in terms of reconstruction quality for this application. Furthermore, the algorithm generalizes well even when trained on a small dataset and is robust to noise. 

The simulated acquisitions in this work used a mono-energetic beam and did not include x-ray scattered radiation, so it remains to be seen how our new reconstruction method will handle these factors. In practice the effect of not modeling the spectrum will likely be minimal as regular filtered backprojection reconstructions also do not account for these physical effects and apply a series of precorrection steps to the projection data instead, such as the beam hardening correction described by \cite{Herman1979}. We foresee using the same approach to extend our method to work in clinical data.

We have shown that the method achieves robust and accurate predictions of breast density, which is an important metric relating to masking and cancer risk. As opposed to current density estimation methods based on mammography and DBT projections, which require assumptions and modeling of the image acquisition process, the use of the images produced by DBToR allows for a direct estimation of the amount of dense tissue present in the volume, resulting in estimates to within 3\%. Accurate determination of breast density opens up further opportunities for personalized risk-based screening. 
As is crucial for an accurate dosimetric estimate, the location in the vertical direction of the dense tissue is also estimated with accuracy, resulting in a state-of-the-art dosimetric estimate. It is known that current model dose estimates introduce an average bias of 30\%, and can misrepresent the actual patient-specific dose by up to 120\% \citep{Dance2005, Sechopoulos2012,Hernandez2015}. In comparison, the results obtained here achieve errors below 20\% with no systematic bias. True patient-specific dosimetry could be used, for the first time, to gather dose registries, especially for screening, ensuring the optimal use of this imaging technology, and allowing for continuous monitoring of dose trends and providing valuable data for additional optimization and development of existing and new imaging technologies.

The main limitation of the current work is that it works on a slice by slice basis rather than on a full 3D volume. With this simplification we were able to concentrate on the network structure rather than on the logistics of handling the enormous datasets needed to train a 3D model.

The logical next step for our algorithm is an extension to fully 3D data instead of 2D slices. From there on, we could extend the model-based parts of the deep learning network instead of starting from precorrected projection data, as in current filtered backprojection methods, by including a polychromatic x-ray spectrum, x-ray scatter, and other relevant factors in the forward model. It could also be valuable to optimize the network output specifically for artificial intelligence reading by training the composite network in an end-to-end fashion. Finally, the reconstruction of a diagnostic-quality volume, for interpretation by radiologists, with the potential for the higher vertical resolution obtained here, would be a valuable improvement for clinical performance.

To conclude, we created a deep learning based reconstruction for DBT that was able to achieve accurate predictions of breast density and from there an accurate calculation of patient specific MGD.

\section*{Acknowledgments}
The research was partially supported by the National Cancer Institute, National Institutes of Health (7R01CA181171) and by the Susan G. Komen Foundation for the Cure (IIR13262248).

\bibliographystyle{model2-names.bst}\biboptions{authoryear}
\bibliography{bibliography}

\end{document}